\documentclass[11pt]{article}
\usepackage{latexsym,amssymb,amsmath,times}
\usepackage{enumitem}
\usepackage{float}

\usepackage{hyperref}
\usepackage[square,sort&compress,comma,numbers]{natbib}

\usepackage{hyperref} 
\makeatletter

\@addtoreset{equation}{section} \makeatother

\input amssym.def
\input amssym.tex

\newcommand{\half}{{{\textstyle\frac{1}{2}}}}
\newcommand{\quarter}{{{\textstyle\frac{1}{4}}}}
\newcommand{\be}{\begin{equation}}
\newcommand{\ee}{\end{equation} }
\newcommand{\beqa}{\begin{eqnarray} }
\newcommand{\eeqa}{\end{eqnarray} }
\newcommand{\ba}{\begin{array}}
\newcommand{\ea}{\end{array}}

\newcommand{\SO}{\mathbf{SO}}

\newcommand{\SL}{\mathbf{SL}}

\newcommand{\mba}{{\mathbf{a}}}
\newcommand{\mbb}{{\mathbf{b}}}
\newcommand{\mbc}{{\mathbf{c}}}

\newcommand{\ODD}{\mathbf{O}(D,D)}

\newcommand{\Ott}{\mathbf{O}(10,10)}

\newcommand{\SLf}{\mathbf{SL}(5)}
\newcommand{\slf}{\mathbf{sl}(5)}
\newcommand{\slt}{\mathbf{sl}(10)}

\newcommand\Tr{{\rm Tr}}

\newcommand\rd{{\rm d}}

\newcommand\cL{{\cal L}}
\newcommand\cM{{\cal M}}
\newcommand\cN{{\cal N}}

\newcommand\cQ{{\cal Q}}
\newcommand\cR{{\cal R}}

\newcommand\hcL{{\hat{\cal L}}}

\newcommand\BP{{\rm \scriptscriptstyle{Berman-Perry}}}

\newcommand\dis{\displaystyle}

\newcommand\seceq{=}

\def\bra{\bar{a}}
\def\brb{\bar{b}}

\def\breta{\bar{\eta}}

\newcommand{\na}{{\nabla}}
\newcommand{\trd}{{\bigtriangledown}}

\begin{document}
\begin{titlepage}
\title{
\vskip 2cm U-geometry\,: $\SLf$
\\~\\}
\author{\sc Jeong-Hyuck Park\quad and\quad Yoonji Suh\quad\quad}
\date{}
\maketitle \vspace{-1.0cm}
\begin{center}
~~~\\
Department of Physics, Sogang University, Mapo-gu,  Seoul 121-742, Korea\\
~\\
\texttt{park@sogang.ac.kr\quad\quad\quad\quad yjsuh@sogang.ac.kr}

~\\
~{}\\
~~~\\~\\~\\
\end{center}
\begin{abstract}
\vskip0.2cm
\noindent
Recently Berman and Perry constructed  a four-dimensional  $\cM$-theory effective action which  manifests  $\SLf$  U-duality. Here we propose an underlying differential geometry of it, under  the name `$\SLf$ U-geometry' which   generalizes    the ordinary  Riemannian geometry in an $\SLf$ compatible manner. We   introduce  a `semi-covariant' derivative that can be converted into  fully covariant derivatives after anti-symmetrizing or contracting the $\SLf$  vector indices appropriately. We also derive  fully covariant scalar and Ricci-like curvatures  which constitute   the effective action as well as  the equation of motion. 
\end{abstract}

{\small
\begin{flushleft}
~~
\end{flushleft}}
\thispagestyle{empty}
\end{titlepage}
\newpage
\tableofcontents 
\section{Introduction and Summary}
Duality is arguably   the most characteristic  feature of string/$\cM$-theory~\cite{Cremmer:1979up,Duff:1990hn,Hull:1994ys}.  
While Riemannian geometry singles out the spacetime metric, $g_{\mu\nu}$, as its only fundamental geometric object,   T-duality in string theory or U-duality in $\cM$-theory   put other form-fields at an equal  footing along with the metric. As a consequence, Riemannian geometry appears  incapable  of  manifesting  the duality,   especially  in the formulations of   low energy effective actions.   Novel differential geometry beyond Riemann  is desirable which treats   the metric and the form-fields equally   as geometric objects, and makes  the covariance apparent  under  not only diffeomorphism but also duality transformations.  \\
\indent Despite of recent progress in various limits,   eleven-dimensional $\cM$-theory remains  still ${\cM}$ysterious, not to mention  its full  U-duality group which  was conjectured  to correspond to  a certain Kac-Moody algebra, or an exceptional generalized geometry called $E_{11}$~\cite{West:2001as,West:2010ev,Rocen:2010bk,West:2011mm}.   Yet,  lower dimensional cases turn out to be  more tractable with smaller U-duality groups~\cite{Cremmer:1979up,Duff:1990hn,Hull:1994ys,Hull:2007zu,Pacheco:2008ps,Berman:2010is,Berman:2011cg,Berman:2011jh,Berman:2011pe,Thompson:2011uw,Coimbra:2011ky,Coimbra:2012af,Berman:2012vc,Berman:2012uy,Hatsuda:2012vm,Bakhmatov:2011ab,Malek:2012pw,Dibitetto:2012rk,Triendl:2009ap,Grana:2012zn,Musaev:2013rq}.  Table~\ref{TABU} summarizes   U-duality groups in various spacetime dimensions. 
\begin{table}[H]
\begin{center}
\begin{tabular}{c|c|c|c|c|c|c}
\hline
Spacetime Dimension~&~${D=1}$~&~${D=2}$~&~${D=3}$~&~${D=4}$~&~${D=5}$~&~${6\leq D\leq8}$~\\
\hline
U-duality Group~&~$\mathbf{SO}(1,1)$~&~$\mathbf{SL}(2)$~&~$\mathbf{SL}(3)\times\mathbf{SL}(2)$~&~$\SLf$~&~$\mathbf{SO}(5,5)$~&~$\mathbf{E}_{D}$~\\
\hline
\end{tabular}
\caption{Finite dimensional U-duality groups in various spacetime dimensions } 
\label{TABU}
\end{center}
\end{table}
In particular,  Berman and Perry managed to construct   $\cM$-theory effective actions which manifest a few  U-duality groups,  firstly for ${D=4}$, $\SLf$~\cite{Berman:2010is}, secondly with Godazgar for ${D=5}$, $\mathbf{SO}(5,5)$~\cite{Berman:2011pe}, and thirdly  with Godazgar and West for 
${D=6}$, $\mathbf{E}_{6}$ as well as ${D=7}$, $\mathbf{E}_{7}$~\cite{Berman:2011jh}.  Their constructed actions were written  in terms of a single object called \textit{generalized metric}  which unifies    a three-form and the Riemannian metric. Further, they are invariant under so-called \textit{generalized diffeomorphism}   which combines   the three-form  gauge symmetry and  the ordinary diffeomorphism.  Yet, the invariance under the generalized diffeomorphism  was not transparent   and had to be checked separately by  direct computations,   since the actions were spelled using  `ordinary' derivatives acting on the generalized metric.  The situation   might be comparable  to the case of writing the Riemannian scalar curvature in terms of a metric and its ordinary derivatives   explicitly, and asking for  its diffeomorphism invariance. \\
\indent It is the purpose of  the present paper  to provide an underlying differential geometry  especially for   the case of  ${D=4}$, $\SLf$  U-duality by Berman and Perry~\cite{Berman:2010is}, under the name, `\textit{U-geometry}'. The  approach we follow is essentially  based on our previous experiences with T-duality~\cite{Jeon:2010rw,Jeon:2011kp,Jeon:2011cn,Jeon:2011vx,Jeon:2011sq,Jeon:2012kd,Jeon:2012hp}  where, in collaboration with  Jeon and Lee, we developed  a stringy differential geometry (or \textit{T-geometry})    for  $\ODD$ T-duality manifest string theory effective actions, called double field theory~\cite{Hull:2009mi,Hull:2009zb,Hohm:2010jy,Hohm:2010pp}. While Hitchin's `generalized geometry' formally combines   tangent and cotangent spaces    giving    a  geometric meaning to the $B$-field~\cite{Gualtieri:2003dx,Hitchin:2004ut,Hitchin:2010qz,Grana:2008yw,Koerber:2010bx,Coimbra:2011nw,Coimbra:2012yy},    double field theory (DFT)  generalizes the generalized geometry one step further, as it   doubles    the spacetime dimensions,  from $D$ to ${D+D}$ (\textit{c.f.~}\cite{Tseytlin:1990nb,Tseytlin:1990va,Siegel:1993xq,Siegel:1993th}) and consequently   manifests  the $\ODD$ T-duality group. Yet,  DFT  is not truly doubled since it is subject to the so called {strong constraint} or \textit{section condition} that  all the fields must live on a $D$-dimensional null hyperplane.   \\
\indent Specifically, through \cite{Jeon:2010rw,Jeon:2011kp,Jeon:2011cn,Jeon:2011vx,Jeon:2011sq,Jeon:2012kd,Jeon:2012hp},   we   introduced    an $\ODD$ T-duality compatible \textit{semi-covariant derivative}~\cite{Jeon:2010rw,Jeon:2011cn}\footnote{For a complementary   alternative  approach we refer to \cite{Hohm:2011si,Hohm:2012mf,Hohm:2012gk} (\textit{c.f.~}\cite{Hohm:2011zr,Hohm:2011dv,Hohm:2010xe,Hohm:2011ex,Hohm:2011cp,Hohm:2011nu}) where a  fully covariant yet  non-physical derivative was  discussed. After projecting out the undecidable  non-physical parts, the two approaches become equivalent.   }. We   extended it to  fermions~\cite{Jeon:2011vx}, to R-R sector~\cite{Jeon:2012kd}, as well as to  Yang-Mills~\cite{Jeon:2011kp}. Then   we constructed, to the  full order in fermions, ten-dimensional supersymmetric double field theories (SDFT)   for  ${\cN=1}$~\cite{Jeon:2011sq} as well as for ${\cN=2}$~\cite{Jeon:2012hp}. Especially the $\cN=2$ ${D=10}$ SDFT  unifies type IIA and IIB supergravities in a  manifestly covariant manner  with respect to     $\Ott$ T-duality and    a `{pair}' of  local Lorentz groups,  besides   the usual general covariance  of  supergravities   or the generalized diffeomorphism. The  distinction of IIA and IIB supergravities  may arise only after  a diagonal gauge fixing of the Lorentz groups:  They are identified as \textit{two different types of solutions} rather than two different theories. \\
\indent For an extension of Hitchin's generalized geometry to $\cM$-theory, we refer to  the works by  
Coimbra, Strickland-Constable and Waldram~\cite{Coimbra:2011ky,Coimbra:2012af} which utilize the  extended tangent space~\cite{Hull:2007zu,Pacheco:2008ps}, but did not make direct connection to the works by Berman and Perry~\cite{Berman:2010is,Berman:2011pe,Berman:2011jh}. \\
\indent The rest of the paper is organized as follows. Below, as for a  convenient quick reference  ---especially  for those who are already  familiar with the works by Berman and Perry--- we summarize  our main results.  For the self-contained systematic  analysis, section~\ref{SECprel} is preliminary. In particular, we   identify  an \textit{integral measure} of the  $\SLf$ U-geometry. In section~\ref{SECcon}, we discuss in detail   the \textit{semi-covariant derivative} as well as  its  full covariantization.  Section~\ref{SECcurv} contains the derivations of \textit{a fully covariant scalar curvature} and \textit{a fully covariant Ricci-like curvature},  which constitute   the  effective action as well as  the equation of motion. In section~\ref{sectionPARA}, U-geometry is reduced to Riemannian geometry. We conclude  with some   comments in section~\ref{SECcomment}. We point out an intriguing connection to  $AdS_{4}$.  \\

\indent\textbf{\large{Summary}}
\begin{itemize}
\item \textit{Notation}: small  Latin alphabet letters denote  the $\SLf$ fundamental indices, as $a,b=1,2,3,4,5$.
\item Assuming   the section condition, $\partial_{[ab}\partial_{cd]}=0$,  we define  a \textit{semi-covariant derivative}~(\ref{semicov1}) and (\ref{Gamma}), relevant for the $\SLf$  covariant generalized Lie derivative, $\hcL_{X}$~(\ref{gLDw}),    (\textit{c.f.~}\cite{Coimbra:2011ky}),
\be
\ba{ll}
\na_{cd}T^{a_{1}a_{2}\cdots a_{p}}{}_{b_{1}b_{2}\cdots b_{q}}:=\!&
\partial_{cd}T^{a_{1}a_{2}\cdots a_{p}}{}_{b_{1}b_{2}\cdots b_{q}}
+\half(\half p-\half q+\omega)\Gamma_{cde}{}^{e}T^{a_{1}a_{2}\cdots a_{p}}{}_{b_{1}b_{2}\cdots b_{q}}\\
{}&~-\sum_{i=1}^{p}T^{a_{1}\cdots e\cdots a_{p}}{}_{b_{1}b_{2}\cdots b_{q}}\Gamma_{cde}{}^{a_{i}}+\sum_{j=1}^{q}\Gamma_{cdb_{j}}{}^{e}T^{a_{1}a_{2}\cdots a_{p}}{}_{b_{1}\cdots e\cdots b_{q}}\,,
\ea
\label{0semicov1}
\ee
where the connection is given  in terms of    an  $\SLf$ generalized metric, $M_{ab}$,  by
\be
\ba{l}
\Gamma_{abc}{}^{d}=\left[B_{[ab]ce}+\half(B_{beac}-B_{aebc}+B_{acbe}-B_{bcae})\right]M^{ed}\,,\\
B_{abcd}=A_{abcd}+\textstyle{\frac{2}{3}}A_{e(ab)}{}^{e}M_{cd}=B_{ab(cd)}\,,\\
A_{abcd}=\half M_{cd}M^{ef}\partial_{ab}M_{ef}-\half\partial_{ab}M_{cd}=A_{[ab](cd)}=B_{[ab]cd}\,.
\ea
\ee
This connection  is \textit{uniquely} determined by requiring the compatibilities with the generalized metric, $\na_{ab}M_{cd}=0$, and with  the generalized Lie derivative, $\hcL_{X}(\partial_{ab})=\hcL_{X}(\na_{ab})$, in addition to a certain `kernel' condition, ${J_{abcd}{}^{efgh}\Gamma_{efgh}=0}$ (\ref{lastcon}).  Generically our  semi-covariant derivative  is not by itself fully covariant, \textit{i.e.~}$\delta_{X}\na_{ab}\neq\hcL_{X}\na_{ab}$, though there are some exceptions (\ref{exc1}\,--\,\ref{exc4}).\\

\item The characteristic feature  of the semi-covariant derivative is that, by  (anti-)symmetrizing or contracting  the $\SLf$ vector indices properly, it can \textit{generate} fully covariant derivatives~(\ref{COD1}\,--\,\ref{COD3}):
\be
\ba{llll}
\na_{[ab}T_{c_{1}c_{2}\cdots c_{q}]}\,,~~~&~~~\na_{ab}T^{a}\,,~~~&~~~
\na^{a}{}_{b}T_{[ca]}+\na^{a}{}_{c}T_{[ba]}\,,~~~&~~~\na^{a}{}_{b}T_{(ca)}-\na^{a}{}_{c}T_{(ba)}\,,\\
\multicolumn{4}{c}{\na_{ab}T^{[abc_{1}c_{2}\cdots c_{q}]}~~~(\mbox{divergences})\,,~~~~~~~~
\na_{ab}\na^{[ab}T^{c_{1}c_{2}\cdots c_{q}]}~~~(\mbox{Laplacians})\,.}
\ea
\ee
~\\

\item While the usual  field strength,  \textit{i.e.~}$R_{abcde}{}^{f}=\partial_{ab}\Gamma_{cde}{}^{f}-\partial_{cd}\Gamma_{abe}{}^{f}+
\Gamma_{abe}{}^{g}\Gamma_{cdg}{}^{f}-\Gamma_{cde}{}^{g}\Gamma_{abg}{}^{f}$,  turns out to be  non-covariant, the following are {\textit{fully covariant}}.
\begin{itemize}
\item $\SLf$ U-geometry \textit{Ricci} curvature~(\ref{RICCI}), 
\be
\ba{l}
\cR_{ab}:=\half R_{(a}{}^{cd}{}_{b)cd}+\half R_{d(a}{}^{cd}{}_{b)c}+
\half\Gamma^{cd}{}_{(a}{}^{e}\Gamma_{b)ecd}-\half\Gamma_{(a}{}^{c\,}{}_{b)}{}^{d}(\Gamma^{e}{}_{cde}+\Gamma^{e}{}_{dec})\\
{}\quad\quad\quad+\quarter\Gamma_{c(a}{}^{cd}\Gamma_{b)de}{}^{e}
+\textstyle{\frac{1}{8}}\Gamma_{acd}{}^{d}\Gamma_{b}{}^{c}{}_{e}{}^{e}\,\,.
\ea
\ee
\item $\SLf$ U-geometry \textit{scalar}  curvature~(\ref{SCALAR}),
\be
\cR:=M^{ab}\cR_{ab}=R_{abc}{}^{abc}+\half\Gamma_{abcd}\Gamma^{cdab}-\half(\Gamma^{c}{}_{acb}+\Gamma^{c}{}_{bac})(\Gamma_{d}{}^{bda}+\Gamma_{d}{}^{abd})\,.
\ee
\end{itemize}

\item The four-dimensional $\SLf$ \textit{U-duality manifest   action} is,  with $M=\det(M_{ab})$, \textit{c.f.~}(\ref{actionR}),
\be
\dis{\int_{\Sigma_{4}}}M^{-1\,}\cR\,.
\ee
Up to  surface integral,  this agrees with the   action obtained by Berman and Perry~\cite{Berman:2010is},  \textit{c.f.~}(\ref{BP1}) and (\ref{BP2}).
\item The \textit{equation of motion} corresponds to the vanishing of an Einstein-like  tensor (\ref{EOM}), 
\be
\cR_{ab}+\half M_{ab}\cR=0\,,
\ee
and hence actually, just like the pure Einstein-Hilbert action,  $\cR_{ab}=0$.

\item From a specific  parameterization of the generalized metric in terms of a  metric, a scalar and  a vector (or its hodge dual three-form potential) in four  dimensions,  \textit{c.f.~}(\ref{PARAMETER}) and (\ref{PARAMETER2}),
\be
\ba{ll}
M_{ab}=\left(\ba{cc}
g_{\mu\nu}/\sqrt{-g}~&~v_{\mu}\\
v_{\nu}~&~\sqrt{-g}(-e^{\phi}+v^{2})\ea\right)\,,\quad&\quad
C_{\lambda\mu\nu}=\textstyle{\frac{1}{\sqrt{-g}\,}}\,\epsilon_{\lambda\mu\nu\rho}v^{\rho}\,,
\ea
\ee
it follows that, 
the  U-geometry  scalar curvature reduces, upon the section condition,  to Riemannian  quantities~(\ref{cRuponsection}),
\be
\cR=e^{-\phi}\left[ R_{g}-\textstyle{\frac{7}{2}}\partial_{\mu}\phi\partial^{\mu}\phi+3\Box\phi+\half e^{-\phi}\left(\trd_{\mu}v^{\mu}\right)^{2}\right]\,,
\ee
and the action  becomes, up to surface integral, as we will see in (\ref{Actionuponsection}) and  (\ref{Fdef}),
\be
\dis{\int_{\Sigma_{4}}}M^{-1\,}\cR=\dis{\int}\rd^{4} x~ e^{-2\phi}\sqrt{-g}\left(R_{g}+\textstyle{\frac{5}{2}}\partial_{\mu}\phi\partial^{\mu}\phi-{\textstyle{\frac{1}{48}}}e^{-\phi}F_{\kappa\lambda\mu\nu}F^{\kappa\lambda\mu\nu}\right) \,.
\ee
\end{itemize}

\section{Section condition, Generalized Lie derivative and Integral measure\label{SECprel}}
The only fundamental object in the $\SLf$ U-geometry we propose   is a $5\times 5$  non-degenerate symmetric matrix, or \textit{generalized metric},
\be
M_{ab}=M_{(ab)}\,.
\ee
Like in the Riemannian   geometry, this with its inverse may be used to  freely  raise or lower the positions of the five-dimensional  $\SLf$ vector  indices,\footnote{\textit{c.f.~}\cite{Coimbra:2012af} where the flat $\SO(5)$ invariant metric was used to raise or lower the indices.  } $a,b,c,\cdots$.\\

The spacetime is formally ten-dimensional with the coordinates carrying  a pair of  anti-symmetric $\SLf$  vector indices,
\be
x^{ab}=x^{[ab]}\,.
\ee
We denote the  derivative by
\be
\partial_{ab}=\partial_{[ab]}=\frac{\partial~~}{\partial x^{ab}}\,,
\ee
such that
\be
\partial_{ab}x^{cd}=\delta_{a}^{~c}\delta_{b}^{~d}-\delta_{a}^{~d}\delta_{b}^{~c}\,.
\ee
However,  the theory is not truly ten-dimensional, as it  is subject to a \textit{section condition}:  All the fields are required to live on a four-dimensional  hyperplane, such that    the $\SLf$ d'Alembertian operator must be trivial~\cite{Berman:2011cg},
\be
\partial_{[ab}\partial_{cd]}=0\,,
\ee
when acting on arbitrary fields, $\Phi$, $\Phi^{\prime}$,  as well as their products, 
\be
\ba{ll}
\partial_{[ab}\partial_{cd]}\Phi=\partial_{[ab}\partial_{c]d}\Phi\seceq 0\,,\quad&\quad
\partial_{[ab}\Phi\partial_{cd]}\Phi^{\prime}=
\half\partial_{[ab}\Phi\partial_{c]d}\Phi^{\prime}-\half\partial_{d[a}\Phi\partial_{bc]}\Phi^{\prime}\seceq 0\,.
\ea
\label{seccon}
\ee
For example, for the generalized metric we have
\be
\ba{ll}
\partial_{[ab}\left(M^{ef}\partial_{c]d}M_{ef}\right)=0\,,~~~~&~~~~
M_{ef}\partial_{[ab}M^{ef}M^{gh}\partial_{c]d}M_{gh}=0\,.
\ea
\label{secM}
\ee
~\\

Generalizing the ordinary Lie derivative, \textit{the $\SLf$ covariant generalized Lie derivative} is defined by~\cite{Berman:2011cg,Coimbra:2012af}
\be
\ba{ll}
\hcL_{X}T^{a_{1}a_{2}\cdots a_{p}}{}_{b_{1}b_{2}\cdots b_{q}}:=\!\!&
\half X^{cd}\partial_{cd}T^{a_{1}a_{2}\cdots a_{p}}{}_{b_{1}b_{2}\cdots b_{q}}
+\half(\half p-\half q+\omega)\partial_{cd}X^{cd}T^{a_{1}a_{2}\cdots a_{p}}{}_{b_{1}b_{2}\cdots b_{q}}\\
{}&-\sum_{i=1}^{p}T^{a_{1}\cdots c\cdots a_{p}}{}_{b_{1}b_{2}\cdots b_{q}}\partial_{cd}X^{a_{i}d}
+\sum_{j=1}^{q}\partial_{b_{j}d}X^{cd}T^{a_{1}a_{2}\cdots a_{p}}{}_{b_{1}\cdots c\cdots b_{q}}\,.
\ea
\label{gLDw}
\ee
Here  we let  the tensor density, $T^{a_{1}a_{2}\cdots a_{p}}{}_{b_{1}b_{2}\cdots b_{q}}$,  have the  \textit{total weight},  $\half p-\half q+\omega$: Each upper or  lower  index contributes to the total weight by $+\half$ or $-\half$ respectively, while $\omega$ denotes any possible  \textit{extra weight} of  the tensor density.\\

It follows from a well-known relation, $\delta\ln(\det K)=\Tr(K^{-1}\delta K)$ which holds for an arbitrary square matrix, $K$, that under the infinitesimal transformation generated  by the $\SLf$ covariant  generalized Lie derivative (\ref{gLDw}) for $\omega=0$, we have 
\be
\ba{l}
\delta_{X}\det(K^{ab})=\half X^{cd}\partial_{cd}\det(K^{ab})+\half\partial_{cd}X^{cd}\det(K^{ab})=\half\partial_{cd}\left[X^{cd}\det(K^{ab})\right]\,,\\
\delta_{X}\det(K^{a}{}_{b})=\half X^{cd}\partial_{cd}\det(K^{a}{}_{b})\,,\\
\delta_{X}\det(K_{ab})=\half X^{cd}\partial_{cd}\det(K_{ab})-\half\partial_{cd}X^{cd}\det(K_{ab})\,.\\
\ea
\label{gLDdet}
\ee
This shows that,  $\det(K^{ab})$, $\det(K^{a}{}_{b})$ and $\det(K_{ab})$ acquire  the extra weights, $\omega=+1$, $\omega=0$ and $\omega=-1$  respectively, while, of course, $p=q=0$.  In particular,  since $\det(M^{ab})$ is    a scalar density with the total weight one  as an  $\SLf$ singlet,   we  naturally \underline{let it   serve as   the \textit{integral measure} of  the $\SLf$ U-geometry.} \\

\section{Covariant derivatives\label{SECcon}}
\subsection{Semi-covariant derivative} 
We  propose  an $\SLf$ \textit{compatible  semi-covariant derivative}, in analogy to the one introduced for $\ODD$ T-duality~\cite{Jeon:2010rw,Jeon:2011cn},\footnote{A similar expression to (\ref{semicov1}) yet with   a different connection first appeared  in \cite{Coimbra:2011ky,Coimbra:2012af} for the case of $p=2$, $q=0$ having  the trivial total weight, $\half p-\half q+\omega=0$.   }
\be
\ba{ll}
\na_{cd}T^{a_{1}a_{2}\cdots a_{p}}{}_{b_{1}b_{2}\cdots b_{q}}:=\!&
\partial_{cd}T^{a_{1}a_{2}\cdots a_{p}}{}_{b_{1}b_{2}\cdots b_{q}}
+\half(\half p-\half q+\omega)\Gamma_{cde}{}^{e}T^{a_{1}a_{2}\cdots a_{p}}{}_{b_{1}b_{2}\cdots b_{q}}\\
{}&~-\sum_{i=1}^{p}T^{a_{1}\cdots e\cdots a_{p}}{}_{b_{1}b_{2}\cdots b_{q}}\Gamma_{cde}{}^{a_{i}}+\sum_{j=1}^{q}\Gamma_{cdb_{j}}{}^{e}T^{a_{1}a_{2}\cdots a_{p}}{}_{b_{1}\cdots e\cdots b_{q}}\,,
\ea
\label{semicov1}
\ee
with  the connection specifically  given by
\be
\ba{l}
\Gamma_{abc}{}^{d}=\left[B_{[ab]ce}+\half(B_{beac}-B_{aebc}+B_{acbe}-B_{bcae})\right]M^{ed}\,,\\
B_{abcd}=A_{abcd}+\textstyle{\frac{2}{3}}A_{e(ab)}{}^{e}M_{cd}=B_{ab(cd)}\,,\\
A_{abcd}=\half M_{cd}M^{ef}\partial_{ab}M_{ef}-\half\partial_{ab}M_{cd}=A_{[ab](cd)}=B_{[ab]cd}\,.
\ea
\label{Gamma}
\ee
As shown below, this connection is the unique solution  to the following five  conditions we  require,
\begin{eqnarray}
&&\Gamma_{abcd}+\Gamma_{abdc}=2A_{abcd}\,,\label{compM}\\
&&\Gamma_{abc}{}^{d}+\Gamma_{bac}{}^{d}=0\,,\label{Gab}\\
&&\Gamma_{abc}{}^{d}+\Gamma_{bca}{}^{d}+\Gamma_{cab}{}^{d}=0\,,\label{Gabc}\\
&&\Gamma_{cab}{}^{c}+\Gamma_{cba}{}^{c}=0\,,\label{Gcc}\\
&&J_{abcd}{}^{efgh}\Gamma_{efgh}=0\,,\label{lastcon}
\end{eqnarray}
where for the last constraint (\ref{lastcon}) we set
\be
J_{abcd}{}^{efgh}:=\half\delta_{[a}^{\,[e}\delta_{b]}^{\,f]}\delta_{[c}^{\,[g}\delta_{d]}^{\,h]}
+\half\delta_{[c}^{\,[e}\delta_{d]}^{\,f]}\delta_{[a}^{\,[g}\delta_{b]}^{\,h]}
+\textstyle{\frac{1}{3}}\delta_{[a}^{\,h}M_{b][c}M^{g[e}\delta_{d]}^{\,f]}
+\textstyle{\frac{1}{3}}\delta_{[c}^{\,h}M_{d][a}M^{g[e}\delta_{b]}^{\,f]}\,.
\label{Jdef}
\ee
The first condition~(\ref{compM}) is equivalent to the generalized metric compatibility,
\be
\ba{lll}
\na_{ab}M_{cd}=0~&~\Longleftrightarrow~&~\Gamma_{ab(cd)}=A_{abcd}\,.
\label{metriccom}
\ea
\ee
The second condition~(\ref{Gab}) is natural, from $\partial_{(ab)}=\na_{(ab)}=0$. The next two relations,  (\ref{Gabc}) and (\ref{Gcc}), are the necessary and sufficient conditions which enable us to replace freely  the ordinary derivatives, $\partial_{cd}$, by the semi-covariant derivatives, $\na_{cd}$,  in the definition of the generalized Lie derivative~(\ref{gLDw}), such that
\be
\ba{ll}
\hcL_{X}T^{a_{1}a_{2}\cdots a_{p}}{}_{b_{1}b_{2}\cdots b_{q}}
{}=&
\half X^{cd}\na_{cd}T^{a_{1}a_{2}\cdots a_{p}}{}_{b_{1}b_{2}\cdots b_{q}}
+\half(\half p-\half q+\omega)\na_{cd}X^{cd}T^{a_{1}a_{2}\cdots a_{p}}{}_{b_{1}b_{2}\cdots b_{q}}\\
{}&~-\sum_{i=1}^{p}T^{a_{1}\cdots c\cdots a_{p}}{}_{b_{1}b_{2}\cdots b_{q}}\na_{cd}X^{a_{i}d}
+\sum_{j=1}^{q}\na_{b_{j}d}X^{cd}T^{a_{1}a_{2}\cdots a_{p}}{}_{b_{1}\cdots c\cdots b_{q}}\,.
\ea
\ee
Eq.(\ref{lastcon}) is the last  condition that   fixes our connection uniquely as spelled in (\ref{Gamma}). We may view the three constraints,   (\ref{Gabc}), (\ref{Gcc}) and (\ref{lastcon}), as the torsionless conditions of the  $\SLf$ U-geometry. \\

It is worthwhile to note that,  the connection satisfies
\be
\ba{l}
\Gamma_{abcd}=A_{abcd}+\Gamma_{[ab][cd]}\,,\\
\Gamma_{abe}{}^{e}=2\Gamma_{eba}{}^{e}=-2\Gamma_{eab}{}^{e}=A_{abe}{}^{e}=2M^{ef}\partial_{ab}M_{ef}\,,
\ea
\label{Gammaprop}
\ee
and,  from (\ref{secM}) due to the section condition,  we have
\be
\ba{ll}
\partial_{[ab}\Gamma_{c]de}{}^{e}=0\,,\quad&\quad
\Gamma_{abe}{}^{e}\Gamma_{cdf}{}^{f}+\Gamma_{bce}{}^{e}\Gamma_{adf}{}^{f}+\Gamma_{cae}{}^{e}\Gamma_{bdf}{}^{f}=0\,.
\ea
\label{GGsec}
\ee
Further, $J_{abcd}{}^{efgh}$ (\ref{Jdef}) satisfies 
\be
\ba{l}
J_{abcd}{}^{efgh}=J_{[ab][cd]}{}^{[ef]gh}=J_{cdab}{}^{efgh}\,,\\
J^{e}{}_{aeb}{}^{klmn}=J^{e}{}_{bea}{}^{klmn}=
\textstyle{\frac{1}{8}}M^{nl}\left(\delta_{a}^{\,m}\delta_{b}^{\,k}+\delta_{b}^{\,m}\delta_{a}^{\,k}
-\textstyle{\frac{2}{3}}M_{ab}M^{km}\right)
-\textstyle{\frac{1}{8}}M^{nk}\left(\delta_{a}^{\,m}\delta_{b}^{\,l}+\delta_{b}^{\,m}\delta_{a}^{\,l}
-\textstyle{\frac{2}{3}}M_{ab}M^{lm}\right),
\ea
\ee
and
\be
J_{abcd}{}^{efgh}J_{efgh}{}^{klmn}=J_{abcd}{}^{klmn}
+\textstyle{\frac{1}{6}}\left(M_{ad}J^{e}{}_{bec}{}^{klmn}
-M_{bd}J^{e}{}_{aec}{}^{klmn}
+M_{bc}J^{e}{}_{aed}{}^{klmn}
-M_{ac}J^{e}{}_{bed}{}^{klmn}\right),
\label{Jsquare}
\ee
which are all consistent with the conditions~(\ref{Gcc}) and (\ref{lastcon}). For example,  the closeness (\ref{Jsquare}) gives  $J_{abcd}{}^{efgh}J_{efgh}{}^{klmn}\Gamma_{klmn}=0$. \\

\underline{The \textit{uniqueness} of the connection can be proven as follows.} \\
First of all, it is straightforward to check that the connection~(\ref{Gamma}) satisfies the five conditions  (\ref{compM}), (\ref{Gab}), (\ref{Gabc}), (\ref{Gcc}), (\ref{lastcon}). We suppose that a generic connection may contain an extra piece, say $\Delta_{abc}{}^{d}$, which we aim to  show trivial. The first four conditions, (\ref{compM}), (\ref{Gab}), (\ref{Gabc}), (\ref{Gcc}) imply 
\begin{eqnarray}
&&\Delta_{abcd}=\Delta_{[ab][cd]}\,,\label{D22}\\
&&\Delta_{[abc]d}=0\,,\label{D3}\\
&&\Delta_{e(ab)}{}^{e}=0\,.\label{Dcc}
\end{eqnarray}
Contacting $a$ and $d$ indices in (\ref{D3}), we further obtain  $\Delta_{e[ab]}{}^{e}=0$. Thus, with  (\ref{Dcc}),   we have
\be
\ba{ll}
\Delta_{eab}{}^{e}=0\,,\quad&\quad\Delta^{e}{}_{aeb}=0\,.
\ea
\ee
The last condition~(\ref{lastcon}) now implies
\be
\Delta_{[ab][cd]}+\Delta_{[cd][ab]} =0\,.
\label{Dabcd}
\ee
Finally, utilizing (\ref{D22}), (\ref{D3}) and (\ref{Dabcd}) fully, we note
\be
\ba{ll}
\Delta_{abcd}=-\Delta_{cdab}=\Delta_{dacb}+\Delta_{acdb}=-\Delta_{bcad}-\Delta_{acbd}=\Delta_{abcd}+2\Delta_{cabd}\,.
\ea
\ee
Therefore, as we aimed, 
\be
\Delta_{cabd}=0\,.
\ee
Namely,   the connection given in (\ref{Gamma}) is the unique connection satisfying the five conditions  (\ref{compM}), (\ref{Gab}), (\ref{Gabc}), (\ref{Gcc}) and (\ref{lastcon}). This completes our proof of the uniqueness.\\

\subsection{Full covariantization} 
Under   the infinitesimal  transformation of the generalized metric,   given in terms of the generalized Lie derivative,
\be
\delta_{X}M_{ab}=\hcL_{X}M_{ab}=\na_{ac}X_{b}{}^{c}+\na_{bc}X_{a}{}^{c}-\half M_{ab}\na_{cd}X^{cd}\,,
\label{deltaM}
\ee
we have 
\be
\delta_{X}A_{abcd}=\hcL_{X}A_{abcd}-\half(\partial_{ab}\partial_{ce}X^{fe})M_{fd}-\half(\partial_{ab}\partial_{de}X^{fe})M_{fc}\,,
\ee
and consequently, 
\be
\delta_{X}\Gamma_{abc}{}^{d}=\hcL_{X}\Gamma_{abc}{}^{d}-\partial_{ab}\partial_{ce}X^{de}+\quarter H_{abc}{}^{d}\,.
\label{deltaXG}
\ee
Here  we set the shorthand notations,  
\be
\ba{l}
H_{abcd}:=I_{abcd}+I_{cdab}-I_{cdba}-I_{abdc}\,,\\
I_{abc}{}^{d}:=\partial_{ab}\partial_{ce}X^{de}
-\textstyle{\frac{1}{3}}M_{ac}\partial^{f}{}_{b}\partial_{fe}X^{de}
+\textstyle{\frac{1}{3}}M_{bc}\partial^{f}{}_{a}\partial_{fe}X^{de}=I_{[ab]c}{}^{d}\,.
\ea
\ee

Before we proceed further, it is worthwhile to analyze  the properties of $H_{abcd}$. Firstly,  it satisfies precisely the same symmetric properties as the standard Riemann curvature, 
\begin{eqnarray}
&&H_{abcd}=H_{[ab][cd]}=H_{cdab}\,, \label{Hab}\\
&&H_{abc}{}^{d}+H_{bca}{}^{d}+H_{cab}{}^{d}=0\,.\label{Habc}
\end{eqnarray}
Secondly,  from
\be
\ba{ll} 
\partial^{e}{}_{a}\partial_{eb}X^{ab}=0\,,\quad&\quad
\partial_{c(a}\partial_{b)d}X^{cd}=0\,.
\ea
\label{ppX}
\ee
it follows that
\be
H_{acb}{}^{c}=0\,.
\label{Hbb}
\ee
Besides, $H_{abcd}$ can be expressed in terms of $J_{abcd}{}^{efgh}$  given in (\ref{Jdef}) as
\be
H_{abcd}=4J_{abcd}{}^{efg}{}_{h\,}\partial_{ef}\partial_{gk}X^{hk}\,,
\label{HJX}
\ee
and hence, with  (\ref{Jsquare}) and (\ref{Hbb}), it further satisfies 
\be
\ba{ll}
H_{abcd}=J_{abcd}{}^{efgh}H_{efgh}\,,\quad&\quad
J_{abc}{}^{befgh}H_{efgh}=H_{abc}{}^{b}=0\,.
\ea
\label{HJH}
\ee
~\\

Now   for an arbitrary  covariant  tensor density, satisfying   
\be
\delta_{X}T^{a_{1}a_{2}\cdots a_{p}}{}_{b_{1}b_{2}\cdots b_{q}}=\hcL_{X}T^{a_{1}a_{2}\cdots a_{p}}{}_{b_{1}b_{2}\cdots b_{q}}\,,
\label{covT}
\ee
straightforward computation may show  
\be
\ba{ll}
\delta_{X}(\na_{ab}T^{c_{1}c_{2}\cdots c_{p}}{}_{d_{1}d_{2}\cdots d_{q}})=&\hcL_{X}\left(\na_{ab}T^{c_{1}c_{2}\cdots c_{p}}{}_{d_{1}d_{2}\cdots d_{q}}\right)\\
{}&
-\quarter\sum_{i=1}^{p}T^{c_{1}\cdots e\cdots c_{p}}{}_{d_{1}d_{2}\cdots d_{q}}H_{abe}{}^{c_{i}}
+\quarter\sum_{j=1}^{q}H_{abd_{j}}{}^{e}T^{c_{1}c_{2}\cdots c_{p}}{}_{d_{1}\cdots e\cdots d_{q}}\,.
\ea
\label{semicovT}
\ee
Hence,  the  semi-covariant derivative of  a generic covariant tensor density  is not necessarily covariant.\\

Yet, for consistency, the metric compatibility of the semi-covariant derivative (\ref{metriccom})  is exceptional,  according to  (\ref{Hab}), 
\be
\ba{ll}
\na_{ab}M_{cd}=0\,,
\quad&\quad
\delta_{X}(\na_{ab}M_{cd})=\hcL_{X}(\na_{ab}M_{cd})=0\,.
\ea
\label{exc1}
\ee
Other exceptional cases include  a scalar density with an arbitrary extra weight,  
\be
\ba{ll}
\na_{ab}\phi=\partial_{ab}\phi+\half\omega\Gamma_{abc}{}^{c}\phi\,,\quad&\quad
\delta_{X}(\na_{ab}\phi)=\hcL_{X}(\na_{ab}\phi)\,,
\ea
\label{exc2}
\ee
the Kronecker delta symbol,
\be
\ba{ll}
\na_{ab}\delta^{c}_{~d}=0\,,\quad&\quad\delta_{X}(\na_{ab}\delta^{c}_{~d})=
\hcL_{X}(\na_{ab}\delta^{c}_{~d})=0\,,
\ea
\label{exc3}
\ee
and, with (\ref{deltaXG}), (\ref{HJX}) and (\ref{HJH}),  the `kernel' condition of the connection,
\be
\ba{ll}
J_{abcd}{}^{efgh}\Gamma_{efgh}=0\,,\quad&\quad
\delta_{X}(J_{abcd}{}^{efgh}\Gamma_{efgh})=\hcL_{X}(
J_{abcd}{}^{efgh}\Gamma_{efgh})=0\,.
\ea
\label{exc4}
\ee
In particular,  from (\ref{exc1}) and (\ref{exc2}), the $\SLf$ U-geometry integral measure, $M^{-1}=\det(M^{ab})$ having  ${\omega=1}$, is covariantly constant,
\be
\na_{ab}M^{-1}=0\,,
\ee
which is also a covariant statement  as
\be
\delta_{X}(\na_{ab}M^{-1})=
\hcL_{X}(\na_{ab}M^{-1})=0\,.
\ee
~\\

The crucial characteristic property of our semi-covariant derivative is that, by  (anti-)symmetrizing or contracting  the $\SLf$ vector  indices appropriately   it may generate fully covariant derivatives:  From (\ref{Habc}) and (\ref{Hbb}), the following  quantities are \textit{fully covariant},
\begin{eqnarray}
&\na_{[ab}T_{c_{1}c_{2}\cdots c_{q}]}\,,&\label{COD1}\\
&\na_{ab}T^{a}\,,&\\
&\na^{a}{}_{b}T_{[ca]}+\na^{a}{}_{c}T_{[ba]}\,,&\label{Tasym}\\
&\na^{a}{}_{b}T_{(ca)}-\na^{a}{}_{c}T_{(ba)}\,,&\label{Tsym}\\
&\na_{ab}T^{[abc_{1}c_{2}\cdots c_{q}]}~~~~~~~:~~~~\mbox{`divergences'}\,,&\label{COD2}\\
&\na_{ab}\na^{[ab}T^{c_{1}c_{2}\cdots c_{q}]}~~:~~~~\mbox{`Laplacians'}\,,&
\label{COD3}
\end{eqnarray}
satisfying  $\delta_{X}(\na_{[ab}T_{c_{1}c_{2}\cdots c_{q}]})= \hcL_{X}(\na_{[ab}T_{c_{1}c_{2}\cdots c_{q}]})$,   $\delta_{X}(\na_{ab}T^{a})=\hcL_{X}(\na_{ab}T^{a})$, \textit{etc.}  Note that the nontrivial values of $q$ in (\ref{COD1}), (\ref{COD2}) and  (\ref{COD3}) are restricted to $q=0,1,2,3$ only, since the anti-symmetrization of more than five $\SLf$ vector  indices is trivial. \\

 Of course, from the metric compatibility, $\na_{ab}M_{cd}=0$~(\ref{metriccom}),   the $\SLf$  indices above may be freely  raised or lowered without breaking the  full covariance: For example, $\na^{[ab}T^{c_{1}c_{2}\cdots c_{q}]}$ is also equally fully covariant along with (\ref{COD1}).\\
 
 Further, in particular, for the case of $q=0$,  the divergence~(\ref{COD2}) reads explicitly,
\be
\na_{ab}T^{ab}=\partial_{ab}T^{ab}+\half(\omega-1)\Gamma_{abc}{}^{c}T^{ab}\,,
\ee
and hence,
\be
\na_{ab}T^{ab}=\partial_{ab}T^{ab}\quad\mbox{for}\quad\omega=1\,,
\ee
which will be  relevant to  `total derivatives' or `surface integral'  in   the effective action.  \\

Successive  applications of the above procedure to a scalar as well as  to  a vector  ---or directly from (\ref{semicov2T})---   lead to  the following second-order  
covariant derivatives,  
 \be
\ba{lll}
\na_{[ab}\na_{cd]}\phi=0\,,\quad&\quad\na_{[ab}\na_{cd}T_{e]}=0\,,\quad&\quad
\na_{[ab}\na_{c]d}T^{d}=0\,,
\ea
\ee
which turn out to be  all \textit{trivial}, \textit{i.e.~}identically vanishing,  due to  (\ref{GGsec}), (\ref{Gab}),  (\ref{Gabc}), (\ref{Gcc}) and  the section condition~(\ref{seccon}).  
Similarly,  for arbitrary  scalar and  vector, we have an identity, 
\be
\na_{[ab}\phi\,\na_{cd}T_{e]}=0\,.
\ee


\section{Curvatures\label{SECcurv}}
The commutator of the $\SLf$ compatible semi-covariant derivatives (\ref{semicov1}) 
leads to  the  following expression,\footnote{In (\ref{nana}), for simplicity, we assume  a trivial extra weight, \textit{i.e.~}${\omega=0}$.} 
\be
\ba{l}
{}\left[\na_{ab},\na_{cd}\right]T^{e_{1}\cdots e_{p}}{}_{f_{1}\cdots f_{q}}\\=
\quarter (p-q) R_{abcdk}{}^{k}T^{e_{1}\cdots e_{p}}{}_{f_{1}\cdots f_{q}}
-\sum_{i}T^{e_{1}\cdots g\cdots e_{p}}{}_{f_{1}\cdots f_{q}}R_{abcdg}{}^{e_{i}}
+\sum_{j}R_{abcdf_{j}}{}^{g}T^{e_{1}\cdots e_{p}}{}_{f_{1}\cdots g\cdots f_{q}}\\
\quad +\left(2\Gamma_{ab[c}{}^{g}\delta_{d]}^{~h}-2\Gamma_{cd[a}{}^{g}\delta_{b]}^{~h}
-\half\Gamma_{abk}{}^{k}\delta_{c}^{~g}\delta_{d}^{~h}+\half\Gamma_{cdk}{}^{k}\delta_{a}^{~g}\delta_{b}^{~h}\right)\na_{gh}T^{e_{1}\cdots e_{p}}{}_{f_{1}\cdots f_{q}}\,,
\ea
\label{nana}
\ee
where $R_{abode}{}^{f}$ denotes the  standard curvature, or the field strength of the connection,
\be
\ba{ll}
R_{abcde}{}^{f}&\!:=\partial_{ab}\Gamma_{cde}{}^{f}-\partial_{cd}\Gamma_{abe}{}^{f}+
\Gamma_{abe}{}^{g}\Gamma_{cdg}{}^{f}-\Gamma_{cde}{}^{g}\Gamma_{abg}{}^{f}\\
{}&=\na_{ab}\Gamma_{cde}{}^{f}+\half\Gamma_{abg}{}^{g}\Gamma_{cde}{}^{f}+\Gamma_{cde}{}^{g}\Gamma_{abg}{}^{f}-\Gamma_{abc}{}^{g}\Gamma_{gde}{}^{f}
-\Gamma_{abd}{}^{g}\Gamma_{cge}{}^{f}\,-\,\left[(a,b)\leftrightarrow(c,d)\right]\,.
\ea
\label{curvatureR}
\ee
Similarly,  straightforward computation shows that the Jacobi identity  reads 
\be
\ba{ll}
0\!&=\Big(\left[\na_{ab},\left[\na_{cd},\na_{ef}\right]\right]+\left[\na_{cd},\left[\na_{ef},\na_{ab}\right]\right]+\left[\na_{ef},\left[\na_{ab},\na_{cd}\right]\right]\Big)T^{g_{1}\cdots g_{p}}{}_{h_{1}\cdots h_{q}}\\
{}&=-\sum_{i}T^{g_{1}\cdots m\cdots g_{p}}{}_{h_{1}\cdots h_{q}}\left(\cQ_{abcdefm}{}^{g_{i}}+\cQ_{cdefabm}{}^{g_{i}}+\cQ_{efabcdm}{}^{g_{i}}\right)\\
{}&\quad\quad+\sum_{j}\left(\cQ_{abcdefh_{j}}{}^{m}+\cQ_{cdefabh_{j}}{}^{m}+\cQ_{efabcdh_{j}}{}^{m}\right)T^{g_{1}\cdots g_{p}}{}_{h_{1}\cdots m\cdots h_{q}}\\
{}&\quad\quad+\quarter(p-q)\left(\cQ_{abcdefm}{}^{m}+\cQ_{cdefabm}{}^{m}+\cQ_{efabcdm}{}^{m}\right)T^{g_{1}\cdots g_{p}}{}_{h_{1}\cdots h_{q}}\,,
\ea
\label{cQcQcQ}
\ee
where we set
\be
\ba{ll}
\cQ_{abcdefg}{}^{h}&:=\na_{ab}R_{cdefg}{}^{h}+\Gamma_{abm}{}^{m}R_{cdefg}{}^{h}+2\Gamma_{ab[c}{}^{m}R_{d]mefg}{}^{h}-2\Gamma_{ab[e}{}^{m}R_{f]mcdg}{}^{h}\\
{}&\,=\partial_{ab}R_{cdefg}{}^{h}-R_{cdefg}{}^{m}\Gamma_{abm}{}^{h}+\Gamma_{abg}{}^{m}R_{cdefm}{}^{h}\\
{}&\,=-\cQ_{abefcdg}{}^{h}\,.
\ea
\ee
Hence, the Jacobi identity implies
\be
\cQ_{abcdefg}{}^{h}+\cQ_{cdefabg}{}^{h}+\cQ_{efabcdg}{}^{h}=0\,.
\label{JcbQ}
\ee
~\\

\indent The curvature satisfies  identities that are rather trivial, 
\be
\ba{ll}R_{abcde}{}^{f}+R_{cdabe}{}^{f}=0\,,~~~&~~~R_{[abcd]e}{}^{f}=0\,.
\ea
\ee
On the other hand, from $\left[\na_{ab},\na_{cd}\right]M_{ef}=0$ and (\ref{Gammaprop}) separately,   nontrivial identities are
\be
\ba{ll}
R_{abcdef}+R_{abcdfe}=\half R_{abcdg}{}^{g} M_{ef}\,,\quad&\quad R_{abcdg}{}^{g}=0\,,
\ea
\label{efsym}
\ee
and hence, combining these two, we note
\be
R_{abcdef}=R_{[ab][cd][ef]}=-R_{[cd][ab][ef]}\,.
\label{Rsympro}
\ee
This implies that  the last line in (\ref{cQcQcQ}) is actually trivial as $\cQ_{abcdefg}{}^{g}=0$, and furthermore  that there exists essentially    \textit{only} one  scalar quantity  one can construct   by contracting the indices of $R_{abcdef}$, which is   $R_{abc}{}^{abc}$. \\

Now we proceed to examine  any covariant properties  of the curvature,   $R_{abcdef}$,  as well as  the scalar,  $R_{abc}{}^{abc}$. Since $\na_{ab}$ is \textit{semi-covariant} rather than \textit{ab initio fully covariant}, we expect it is also in a way semi-covariant,  which is also the  case with  T-geometry for double field theory~\cite{Jeon:2011cn}.  In fact,  we shall see  shortly that $R_{abc}{}^{abc}$ and hence $R_{abcdef}$ are not fully covariant, but they provide  building blocks to construct   fully covariant quantities which we shall call fully covariant curvatures. \\

Under the transformation  of the generalized metric set  by the generalized diffeomorphism,   the connection varies as (\ref{deltaXG}),
\be
\delta_{X}\Gamma_{abc}{}^{d}=\hcL_{X}\Gamma_{abc}{}^{d}-\partial_{ab}\partial_{ce}X^{de}+\quarter H_{abc}{}^{d}\,,
\label{Rvar1}
\ee
while the section condition~(\ref{seccon}) implies
\be
\ba{l}
\partial_{ab}\partial_{cd}X^{cd}=2\partial_{ac}\partial_{bd}X^{cd}\,,\\
\partial_{ab}\partial_{ch}X^{gh}\Gamma_{gd(ef)}+\partial_{ab}\partial_{dh}X^{gh}\Gamma_{cg(ef)}-\half\partial_{ab}\partial_{gh}X^{gh}\Gamma_{cd(ef)}=\half\partial_{ab}\partial_{cd}X^{gh}\Gamma_{gh(ef)}\,.
\ea
\label{Rvar2}
\ee
Using  the formulae above, it is straightforward to compute the  variation of the curvature,
\be
\ba{ll}
\delta_{X}R_{abcdef}-\hcL_{X}R_{abcdef}=&\quarter\left(\na_{ab}H_{cdef}+
\half\Gamma_{abg}{}^{g}H_{cdef}-\Gamma_{abc}{}^{g}H_{gdef}-\Gamma_{abd}{}^{g}H_{cgef}\right)\\
{}&+\partial_{ab}\partial_{ch}X^{gh}\Gamma_{gd[ef]}+\partial_{ab}\partial_{dh}X^{gh}\Gamma_{cg[ef]}
-\half\partial_{ab}\partial_{gh}X^{gh}\Gamma_{cd[ef]}\\
{}&\,-\,\left[(a,b)\leftrightarrow(c,d)\right]\,.
\ea
\label{varR}
\ee
As expected, $R_{abcdef}$ itself is not fully covariant. Yet, for consistency, the trivial  quantity,  ${R_{abcd(ef)}=0}$, is fully covariant,  since $H_{ab(cd)}=0$ from (\ref{Hab}).\\

In order to identify  nontrivial fully  covariant curvatures, from (\ref{Rvar1}), we replace $\partial_{ab}\partial_{ce}X^{de}$ in (\ref{varR}) by 
\be
\partial_{ab}\partial_{ce}X^{de}=-(\delta_{X}-\hcL_{X})\Gamma_{abc}{}^{d}
+\quarter H_{abc}{}^{d}\,,
\ee
and using  (\ref{Gammaprop}), (\ref{Hab}), (\ref{Habc}), (\ref{Hbb}), (\ref{usefulGGd}) and (\ref{usefulGG2to1}), we may  organize    the anomalous part in     the  variation of the scalar, $R_{abc}{}^{abc}$, as
\be
(\delta_{X}-\hcL_{X})R_{abc}{}^{abc}
=-(\delta_{X}-\hcL_{X})\Big(\half\Gamma_{abcd}\Gamma^{cdab}-\half\Gamma^{c}{}_{acb}\Gamma^{db}{}_{d}{}^{a}+\half\Gamma_{abc}{}^{c}\Gamma_{d}{}^{adb}+\textstyle{\frac{1}{8}}\Gamma_{abc}{}^{c}\Gamma^{ab}{}_{d}{}^{d}\Big)\,.
\label{varscalarR}
\ee
Therefore,  the following quantity is a genuine fully covariant \underline{\textit{scalar curvature of $\SLf$ U-geometry}},  (\textit{c.f.~}\cite{Coimbra:2012af}),
\be
\ba{ll}
\cR&:=R_{abc}{}^{abc}+\half\Gamma_{abcd}\Gamma^{cdab}-\half(\Gamma^{c}{}_{acb}+\Gamma^{c}{}_{bac})(\Gamma_{d}{}^{bda}+\Gamma_{d}{}^{abd})\\
{}&\,=R_{abc}{}^{abc}+\half\Gamma_{abcd}\Gamma^{cdab}-\half\Gamma^{c}{}_{acb}\Gamma^{db}{}_{d}{}^{a}+\half\Gamma_{abc}{}^{c}\Gamma_{d}{}^{adb}+\textstyle{\frac{1}{8}}\Gamma_{abc}{}^{c}\Gamma^{ab}{}_{d}{}^{d}\,,
\ea
\label{SCALAR}
\ee
satisfying with ${\omega=0}$,
\be
\delta_{X}\cR=\hcL_{X}\cR=\half X^{ab}\partial_{ab}\cR\,.
\ee
~\\

Further, under arbitrary variation of the generalized metric, $\delta M_{ab}$,  the connection transforms as
\be
\ba{l}
\delta A_{abcd}=-\half\na_{ab}\delta M_{cd}+\half M_{cd}M^{ef}\na_{ab}\delta M_{ef}+\Gamma_{ab(c}{}^{e}\delta M_{d)e}\,,\\
\delta \Gamma_{abcd}=\delta(\Gamma_{abc}{}^{e}M_{ed})=
\delta B_{[ab]cd}+\half (\delta B_{bdac}-\delta B_{adbc}+\delta B_{acbd}-\delta B_{bcad})\,,
\ea
\label{deltaA}
\ee
which  induces 
\be
\delta R_{abcde}{}^{f}=\na_{ab}\delta\Gamma_{cde}{}^{f}+\half\Gamma_{abg}{}^{g}\delta\Gamma_{cde}{}^{f}
-\Gamma_{abc}{}^{g}\delta\Gamma_{gde}{}^{f}
-\Gamma_{abd}{}^{g}\delta\Gamma_{cge}{}^{f}\,-\,\left[(a,b)\leftrightarrow(c,d)\right]\,.
\label{deltaR6}
\ee
Now, from (\ref{deltaR6}) alone ---without referring to the details of (\ref{deltaA})--- we may  be able to derive the transformation of the fully covariant scalar curvature as follows\footnote{This is  analogue to the variation of the Riemannian scalar curvature,
\[\delta R=\delta g^{\mu\nu}R_{\mu\nu}+\na_{\mu}\left(g^{\nu\rho}
\delta\Gamma_{\nu\rho}^{\mu}-g^{\mu\nu}\delta\Gamma^{\rho}_{\rho\nu}\right)\,.
\]}
\be
\delta\cR=2\delta M^{ab}\,\cR_{ab}+\na^{ab}\!\left(M_{bc}M^{de}\delta\Gamma_{ade}{}^{c}-\half\delta\Gamma_{abc}{}^{c}\right)\,,
\label{deltacR}
\ee
which in turn gives rise to the following  fully covariant  \underline{\textit{Ricci curvature of  $\SLf$ U-geometry}},  (\textit{c.f.~}\cite{Coimbra:2012af}),
\be
\cR_{ab}:=\half R_{(a}{}^{cd}{}_{b)cd}+\half R_{d(a}{}^{cd}{}_{b)c}+
\half\Gamma^{cd}{}_{(a}{}^{e}\Gamma_{b)ecd}-\half\Gamma_{(a}{}^{c\,}{}_{b)}{}^{d}(\Gamma^{e}{}_{cde}+\Gamma^{e}{}_{dec})+\quarter\Gamma_{c(a}{}^{cd}\Gamma_{b)de}{}^{e}
+\textstyle{\frac{1}{8}}\Gamma_{acd}{}^{d}\Gamma_{b}{}^{c}{}_{e}{}^{e}\,,
\label{RICCI}
\ee
satisfying  
\be
\ba{lll}
\cR_{ab}=\cR_{ba}\,,\quad&\quad M^{ab}\cR_{ab}=\cR\,,
\ea
\label{cRabprop}
\ee
and
\be
\delta_{X}\cR_{ab}=\hcL_{X}\cR_{ab}\,.
\label{COVRicci}
\ee
~\\
\indent  Naturally, the four-dimensional $\SLf$ U-duality manifest  effective action reads
\be
\dis{\int_{\Sigma_{4}}}M^{-1\,}\cR\,,
\label{actionR}
\ee
where $\Sigma_{4}$ denotes the four-dimensional hyperplane where the theory lives  to satisfy the section condition~(\ref{seccon}). As shown through (\ref{BP1}) and (\ref{BP2}) in  Appendix~\ref{SECBP}, up to surface integral,   this action agrees with the   action obtained  by Berman and Perry~\cite{Berman:2010is}.\\
 
From (\ref{deltacR}),   the action transforms under arbitrary variation of the generalized metric, 
\be
\delta\left(\dis{\int_{\Sigma_{4}}}M^{-1\,}\cR\right)=
\dis{\int_{\Sigma_{4}}}M^{-1\,}\delta M^{ab}(2\cR_{ab}+M_{ab}\cR)\,.
\ee
Hence, the \textit{equation of motion} corresponds to  the vanishing of  the following Einstein-like tensor,\footnote{Note the plus sign in (\ref{EOM}) in comparison to the Riemannian Einstein tensor, $R_{\mu\nu}-\half g_{\mu\nu}R$.}
\be
\cR_{ab}+\half M_{ab}\cR=0\,,
\label{EOM}
\ee
and hence, it follows
\be
\cR_{ab}=0\,.
\label{EOM2}
\ee
This also (indirectly) verifies   the covariance of the Ricci-like curvature~(\ref{COVRicci}), since any symmetry of the action ---in this case the generalized diffeomorphism--- is also a symmetry of the equation of motion.\footnote{As discussed in section~\ref{sectionPARA}, upon the section condition the U-geometry action~(\ref{actionR})  reduces to a familiar   Riemannian action~(\ref{Actionuponsection})  of which the   equations motion, \textit{c.f.~}(\ref{EOM2}),  are surely fully covariant. See also \textit{e.g.~}\cite{Bekaert:2009zz} for general analysis and proof.} Further, from the invariance of the action under the generalized diffeomorphism~(\ref{deltaM}), a conservation relation follows
\be
\na^{c}{}_{[a}\cR_{b]c}+\textstyle{\frac{3}{8}}\na_{ab}\cR=0\,,
\label{conservation}
\ee
which may be also directly verified using \textit{e.g.~}(\ref{JcbQ}). \\



\section{Parametrization and  Reduction to Riemann \label{sectionPARA}}
We  parametrize  the generalized metric, \textit{i.e.~}a generic non-degenerate $5\times 5$ symmetric matrix,   by
\be
M_{ab}=\left(\ba{cc}
g_{\mu\nu}/\sqrt{-g}~&~v_{\mu}\\
v_{\nu}~&~\sqrt{-g}(-e^{\phi}+v^{2})\ea\right)\,,
\label{PARAMETER}
\ee
where $\phi$, $v^{\mu}$ and $g_{\mu\nu}$ denote a scalar, a vector and  a Riemannian   metric  in Minkowskian four-dimensions, such that $v_{\mu}=g_{\mu\nu}v^{\nu}$, $v^{2}=g^{\mu\nu}v_{\mu}v_{\nu}$  and $g=\det(g_{\mu\nu})$.  
The vector can be  dualized   to a three-form,\footnote{In our convention, $\epsilon^{0123}=1$.  }
\be
C_{\lambda\mu\nu}=\textstyle{\frac{1}{\sqrt{-g}\,}}\epsilon_{\lambda\mu\nu\rho}v^{\rho}\,,
\label{PARAMETER2}
\ee 
which  may couple to a membrane.  \\

The existence of the scalar might appear odd especially if the spacetime dimension were eleven rather than four. However, without  the scalar, the (off-shell) degrees of freedom would not  match in  the above decomposition of the generalized metric,   
\be
15\,=\,1\,+\,4\,+\,10\,\neq\,4\,+\,10\,.
\label{Match}
\ee
Moreover,  with a parametrization of an $\slf$ Lie algebra element, \textit{i.e.}~a generic  $5\times 5$ traceless matrix,
\be
H_{a}{}^{b}=\left(\ba{cc}\mba_{\mu}{}^{\nu}&\,\mbb_{\mu}\\
\mbc^{\nu}&-\mba_{\lambda}{}^{\lambda}\ea\right)\,,
\label{PARAMETER3}
\ee
the infinitesimal  $\slf$ U-duality transformation, $\delta M_{ab}=H_{a}{}^{c}M_{cb}+H_{b}{}^{c}M_{ac}$\,,  amounts to\footnote{In (\ref{infslf}),   the four-dimensional Greek letter indices are raised or lowered by the Riemannian metric from the default positions in (\ref{PARAMETER3}), for example\, $\mba_{\mu\nu}=\mba_{\mu}{}^{\lambda}g_{\lambda\nu}$. }
\be
\ba{l}
\delta\phi=-\left(\mba_{\lambda}{}^{\lambda}+\sqrt{-g}\,\mbb_{\lambda}v^{\lambda}\right)\,,\\
\delta v_{\mu}=\mba_{\mu}{}^{\lambda}v_{\lambda}-\mba_{\lambda}{}^{\lambda}v_{\mu}+\sqrt{-g}\left(-e^{\phi}+v^{2}\right)\mbb_{\mu}+\frac{1}{\sqrt{-g}}\,\mbc_{\mu}\,,\\
\delta g_{\mu\nu}=\mba_{\mu\nu}+\mba_{\nu\mu}-\mba_{\lambda}{}^{\lambda}g_{\mu\nu}+\sqrt{-g}\left(\mbb_{\mu}v_{\nu}+\mbb_{\nu}v_{\mu}-\mbb_{\lambda}v^{\lambda}g_{\mu\nu}\right)\,.
\ea
\label{infslf}
\ee 
Clearly  this confirms that    the scalar   is inevitable     for the    closeness of  the  U-duality  transformations:  Setting $\mba_{\lambda}{}^{\lambda}\equiv 0$ and $\mbb_{\lambda}\equiv 0$ for  $\delta\phi\equiv0$ would break  the $\SLf$ U-duality group  to its subgroup,  $\mathbf{R}^{4}\rtimes\mathbf{SL}(4)$.\\

Similarly, under  the infinitesimal transformation set by the generalized Lie derivative~(\ref{deltaM}),  with the parameter $(X^{\mu\nu},X^{\mu 5})=(\half\epsilon^{\mu\nu\rho\sigma}\Lambda_{\rho\sigma},\xi^{\mu})$ and 
upon the choice of the  `section' by   $({\partial_{\mu\nu},\partial_{\mu 5})\equiv (0,\partial_{\mu})}$,  each component field transforms as  (\textit{c.f.~}\cite{Berman:2011cg})
\be
\ba{l}
\delta\phi=\xi^{\lambda}\partial_{\lambda}\phi=\cL_{\xi}\phi\,,\\
\delta v_{\mu}=\xi^{\lambda}\partial_{\lambda}v_{\mu}+\partial_{\mu}\xi^{\lambda}v_{\lambda}-\frac{1}{\,2\sqrt{-g}}\epsilon_{\mu}{}^{\rho\sigma\tau}\partial_{\rho}\Lambda_{\sigma\tau}=\cL_{\xi}v_{\mu}-\frac{1}{\,2\sqrt{-g}}\epsilon_{\mu}{}^{\rho\sigma\tau}\partial_{\rho}\Lambda_{\sigma\tau}\,,\\
\delta g_{\mu\nu}=\xi^{\lambda}\partial_{\lambda}g_{\mu\nu}+\partial_{\mu}\xi^{\lambda}g_{\lambda\nu}+\partial_{\nu}\xi^{\lambda}g_{\mu\lambda}=\cL_{\xi}g_{\mu\nu}\,.
\ea
\ee
In particular, as expected, the covariant divergence of the vector is a scalar,\footnote{
With the Bianchi identity of the Riemann curvature, 
\[
\trd_{\mu}\left(\frac{1}{\sqrt{-g}}\epsilon^{\mu\rho\sigma\tau}\partial_{\rho}\Lambda_{\sigma\tau}\right)=\frac{1}{2\sqrt{-g}}\epsilon^{\mu\rho\sigma\tau}\left[\trd_{\mu},\trd_{\rho}\right]\Lambda_{\sigma\tau}=
\frac{1}{2\sqrt{-g}}\epsilon^{\mu\rho\sigma\tau}\left(-R^{\lambda}{}_{\sigma\mu\rho}\Lambda_{\lambda\tau}-R^{\lambda}{}_{\tau\mu\rho}\Lambda_{\sigma\lambda}\right)
=0\,.
\]}
$\delta(\trd_{\mu}v^{\mu})=\xi^{\lambda}\partial_{\lambda}(\trd_{\mu}v^{\mu})$.\\

The inverse of the generalized metric and their determinants are 
\be
\ba{ll}
\multicolumn{2}{c}{
M^{ab}=\left(\ba{cc}
\sqrt{-g}(g^{\mu\nu}-e^{-\phi}v^{\mu}v^{\nu})~&~e^{-\phi}v^{\mu}\\
e^{-\phi}v^{\nu}~&~-e^{-\phi}/\sqrt{-g}\ea\right)\,,}\\
\det(M_{ab})=e^{\phi}/\sqrt{-g}\,,~~~~&~~~~
\det(M^{ab})=e^{-\phi}\sqrt{-g}\,,
\ea
\ee
which are consistent with   (\ref{gLDdet}), and in particular   assures us   that  ${M^{-1}=\det(M^{ab})}$  corresponds to the  \textit{$\SLf$  invariant  measure} of the U-geometry.\\

The fully covariant scalar curvature~(\ref{SCALAR}) now  reduces to   Riemannian quantities,  
\be
\cR=e^{-\phi}\left[ R_{g}-\textstyle{\frac{7}{2}}\partial_{\mu}\phi\partial^{\mu}\phi+3\Box\phi+\half e^{-\phi}\left(\trd_{\mu}v^{\mu}\right)^{2}\right]\,,
\label{cRuponsection}
\ee
and hence the action~(\ref{actionR}) becomes, up to surface integral,
\be
\dis{\int_{\Sigma_{4}}}M^{-1\,}\cR=\dis{\int}\rd^{4} x~ e^{-2\phi}\sqrt{-g}\left(R_{g}+\textstyle{\frac{5}{2}}\partial_{\mu}\phi\partial^{\mu}\phi-{\textstyle{\frac{1}{48}}}e^{-\phi}F_{\kappa\lambda\mu\nu}F^{\kappa\lambda\mu\nu}\right) \,,
\label{Actionuponsection}
\ee
where $F_{\kappa\lambda\mu\nu}$ is the field strength of the three-form potential,
\be
F_{\kappa\lambda\mu\nu}=4\partial_{[\kappa}C_{\lambda\mu\nu]}\,.
\label{Fdef}
\ee

\section{Comments\label{SECcomment}}
Like in double field theories (bosonic DFT~\cite{Jeon:2011cn}, $\cN=1$ SDFT~\cite{Jeon:2011sq} and  ${\cN=2}$ SDFT~\cite{Jeon:2012hp}),  according to  (\ref{cRabprop}) and (\ref{EOM2}),   the U-geometry  Lagrangian vanishes on-shell  strictly,  $M^{-1}\cR=0$.  However, this does not necessarily mean that the Riemannian action~(\ref{Actionuponsection}) is trivial, as the difference is  given by  a   nontrivial surface integral.   Hence, in contrast to Riemannian geometry, U-geometry as well as  T-geometry appear to clearly distinguish the bulk Lagrangians  from  the York-Gibbons-Hawking type boundary terms~\cite{York:1972sj,Gibbons:1976ue}, by removing their ambiguity,~\textit{c.f.~}\cite{Berman:2011kg}.\\

In fact, the parametrization of the generalized metric~(\ref{PARAMETER}) we have  considered above possesses the spacetime signature, `$2+3$', \textit{e.g.}~as seen from 
\be
\ba{lll}
M_{ab}=E_{a}{}^{\bra}E_{b}{}^{\brb}\breta_{\bra\brb}\,,~~~~&~~~~
E_{a}{}^{\bra}=\left(\ba{cc}e_{\mu}{}^{i}/\sqrt{e}\,&\,0\\
\sqrt{e}\,v^{\nu}e_{\nu}{}^{i}\,&\,\sqrt{e}\,e^{\phi/2}\ea\right)\,,~~~~&~~~~\breta=\mbox{diag}(-+++-)\,.
\ea
\label{VielbeinE}
\ee
Alternatively, if we had assumed the Minkowskian signature with $\breta=\mbox{diag}(-++++)$, such that $\phi$ had been replaced by $\phi+i\pi$ or $e^{\phi}\rightarrow-e^{\phi}$, the kinetic term of the four-form field strength in the resulting action~(\ref{Actionuponsection}) would have  carried the opposite wrong sign to  break   the unitarity.  Therefore, we conclude that the spacetime signature of the generalized metric ought to  be $2+3$, and  the relevant  internal local Lorentz group should be   $\mathbf{O}(2,3)$. This seems to point to the four-dimensional anti-de Sitter space, $AdS_{4}$.\\ 

 It is desirable to verify (\ref{COVRicci}) and (\ref{conservation}) directly in a covariant manner, for which one might  need more  identities for the curvature in addition to (\ref{Rsympro}) and (\ref{conservation}). \\

Supersymmetrization, reduction to double field theory (\textit{c.f.}~\cite{Thompson:2011uw}) and  extensions to other U-duality groups (\textit{c.f.~}\cite{Coimbra:2011ky,Coimbra:2012af}), especially $E_{11}$~\cite{West:2001as,West:2010ev,West:2011mm},  are of interest for future works.  It is intriguing to note that, the $\SLf$ U-duality group naturally gets embedded into $\SL(10)$ (see \cite{Berman:2010is} and also our Appenix~\ref{SECBP}), which may well hint at higher dimensional larger U-duality groups.\\

\section*{Acknowledgements} 
We wish to thank David Berman  for the kind explanation of his works  during \href{http://cquest-eufp2012.sogang.ac.kr}{\textit{CQUeST EU-FP Workshop}}, \textit{Seoul, 2012}.  JHP also benefits from discussions with Bernard Julia during an Isaac newton Institute 2012 Program, \href{http://sms.cam.ac.uk/collection/1202251}{\textit{Mathematics and Applications of Branes in String and M-theory}}, and also with  Pei-Wen Kao. \\
The work was supported by the National Research Foundation of Korea and  the Ministry of Education, Science and Technology with the Grant   No. 2012R1A2A2A02046739,   No. 2012R1A6A3A03040350, No.  2010-0002980 and  No.  2005-0049409 (CQUeST).  We thank Chris Blair and Emanuel Malek for pointing out   numerical errors in    (\ref{cRuponsection}), (\ref{Actionuponsection}) from  the previous arXiv version. 



\appendix
\begin{center}
{\textbf{\Large{\underline{Appendices\, A \& B}}}}
\end{center}


\section{$\SL(5)\subset\SL(10)$\label{SECBP}}
As a shorthand notation~\cite{Berman:2010is}, we let the capital letters, $A,B,C,\cdots$ represent   pairwise skew-symmetric $\SLf$ indices, such that for the derivative,
\be
\partial_{A}\equiv\partial_{a_{1}a_{2}}\,,
\ee
and for tensors carrying  pairwise skew-symmetry indices,
\be
T^{A_{1}A_{2}\cdots A_{m}}{}_{B_{1}B_{2}\cdots B_{n}}\equiv T^{[a_{11}a_{12}][a_{21}a_{22}]\cdots [a_{m1}a_{m2}]}{}_{[b_{11}b_{12}][b_{21}b_{22}]\cdots [b_{n1}a_{n2}]}\,.
\ee
Being ten-dimensional, the capital letters are essentially  for $\SL(10)$, as the $\slf$ infinitesimal transformation, $w^{a}{}_{b}$ with $w^{a}{}_{a}=0$, acts now as an $\slt$ element: 
\be
\ba{ll}
w^{A}{}_{B}=w^{a_{1}}{}_{[b_{1}}\delta^{a_{2}}{}_{b_{2}]}+\delta^{a_{1}}{}_{[b_{1}}w^{a_{2}}{}_{b_{2}]}\,,~~~~&~~~~w^{A}{}_{A}=0\,.
\ea
\ee
We may further set a \textit{generalized metric for the $\SL(10)$ indices,}
\be
M_{AB}=M_{[a_{1}a_{2}][b_{1}b_{2}]}:=\half(M_{a_{1}b_{1}}M_{a_{2}b_{2}}-M_{a_{1}b_{2}}M_{a_{2}b_{1}})\,.
\ee
It follows that,  the  inverse is given by
\be
M^{AB}=M^{[a_{1}a_{2}][b_{1}b_{2}]}=\half(M^{a_{1}b_{1}}M^{a_{2}b_{2}}-M^{a_{1}b_{2}}M^{a_{2}b_{1}})\,,
\ee
satisfying
\be
M_{AB}M^{BC}=\delta_{A}{}^{C}=\delta_{[a_{1}}^{~[c_{1}}\delta_{a_{2}]}^{~c_{2}]}=\half(\delta_{a_{1}}^{~c_{1}}\delta_{a_{2}}^{~c_{2}}-\delta_{a_{1}}^{~c_{2}}\delta_{a_{2}}^{~c_{1}})\,,
\ee
and the determinant reads
\be
\det (M_{AB})=(\half)^{10}\left[\det(M_{ab})\right]^{4}\,.
\ee 
Henceforth, we  use $M^{AB}$ and $M_{AB}$  to raise and lower the $\slt$ capital letter indices.\\

For (\ref{Gamma}),
\be
A_{abcd}=\half M_{cd}M^{ef}\partial_{ab}M_{ef}-\half\partial_{ab}M_{cd}\,,
\ee
we further set
\be
A_{AB}{}^{C}:=2A_{a_{1}a_{2}[b_{1}}{}^{[c_{1}}\delta_{b_{2}]}^{~c_{2}]}=\quarter\delta_{B}^{~C}(M^{DE}\partial_{A}M_{DE})-\half(\partial_{A}M_{BD})M^{CD}\,,
\label{Acon}
\ee
such that
\be
A_{ABC}=A_{ACB}=\quarter M_{BC}(M^{DE}\partial_{A}M_{DE})-\half\partial_{A}M_{BC}\,,
\label{Asym}
\ee
and
\be
A_{AB}{}^{B}=2M^{DE}\partial_{A}M_{DE}=4A_{a_{1}a_{2}b}{}^{b}=4\Gamma_{a_{1}a_{2}b}{}^{b}=
8M^{bc}\partial_{a_{1}a_{2}}M_{bc}\,.
\ee
~\\

Now, we are ready to compare  our action~(\ref{actionR}) with   the action by Berman and Perry which  was written in terms of  the $\SL(10)$ notation. Up to total derivatives, our scalar curvature~(\ref{SCALAR}) agrees with the Lagrangian by Berman and Perry~\cite{Berman:2010is} as
\be
\cR= \na_{ab}(\Gamma_{c}{}^{abc}-\Gamma_{c}{}^{acb})-\half R_{\BP}\,,
\label{BP1}
\ee
where
\be
\ba{ll}
R_{\BP}\!\!\!&=\textstyle{\frac{1}{12}}M^{ST}\partial_{S}M^{PQ}\partial_{T}M_{PQ}-\half M^{ST}\partial_{S}M^{PQ}\partial_{P}M_{TQ}\\
{}&~\quad+\textstyle{\frac{1}{4}}M^{MN}M^{ST}\partial_{M}M_{NT}(M^{PQ}\partial_{S}M_{PQ})+\textstyle{\frac{1}{12}}M^{ST}(M^{MN}\partial_{S}M_{MN})(M^{PQ}\partial_{T}M_{PQ})\\
{}&=-\textstyle{\frac{1}{3}} A_{ABC}A^{ABC}+2 A_{ABC}A^{BAC}-\textstyle{\frac{3}{4}} A_{AC}{}^{C}A_{D}{}^{DA}+\textstyle{\frac{11}{96}}A_{AC}{}^{C}A^{AD}{}_{D}\\
{}&=-A_{abcd}A^{abcd}+4A_{abcd}A^{acbd}+{\textstyle\frac{3}{2}}A_{abc}{}^{c}A^{ab}{}_{d}{}^{d}
+6A_{abc}{}^{c}A_{d}{}^{abd}-4A_{cab}{}^{c}A_{d}{}^{bad}\,.
\ea
\label{BP2}
\ee
Note also
\be
{\cal R}=-\partial_{ab}(2A^{cab}{}_{c}+A^{abc}{}_{c})+\half A_{abcd}A^{abcd}-2A_{abcd}A^{acbd}-\textstyle{\frac{1}{4}} A_{abc}{}^{c}A^{abd}{}_{d}-2A_{cab}{}^{c}A^{abd}{}_{d}+2A_{cab}{}^{c}A^{dba}{}_{d}\,.
\ee

The remaining of this Appendix is devoted to the construction of another  semi-covariant derivative  which is for the group  $\SL(10)$ and is different from  the one in (\ref{semicov1}) for $\SLf$.  The alternative semi-covariant derivative is defined by employing  $A_{AB}{}^{C}$ (\ref{Acon}) as the connection,  
\be
\ba{ll}
D_{A}T^{B_{1}\cdots B_{m}}{}_{C_{1}\cdots C_{n}}:=&\partial_{A}T^{B_{1}\cdots B_{m}}{}_{C_{1}\cdots C_{n}}+\textstyle{\frac{1}{8}}(m-n)A_{AD}{}^{D}T^{B_{1}\cdots B_{m}}{}_{C_{1}\cdots C_{n}}\\
{}&\,-\sum_{i}T^{B_{1}\cdots D\cdots B_{m}}{}_{C_{1}\cdots C_{n}}A_{AD}{}^{B_{i}}+\sum_{j}
A_{AC_{j}}{}^{E}T^{B_{1}\cdots B_{m}}{}_{C_{1}\cdots E\cdots C_{n}}\,.
\ea
\label{semicov2}
\ee
In contrast to  (\ref{Acon}),   for the  connection of  $\Gamma_{abc}{}^{d}$ defined in   (\ref{Gamma}),    an analogue expression, $\Gamma_{AB}{}^{C}:=\Gamma_{a_{1}a_{2}[b_{1}}{}^{[c_{1}}\delta_{b_{2}]}^{~c_{2}]}$,  cannot  be  written entirely  in a $\SL(10)$ covariant manner, \textit{i.e.~}in terms of $\partial_{A}$ and $M_{AB}$ carrying   the $\SL(10)$  indices  only.\footnote{One might try to look for other connection alternative to the one we constructed    in (\ref{Gamma}), by  \textit{e.g.~}modifying  the index-eight tensor, $J_{abcd}{}^{efgh}$ (\ref{Jdef}),  ---used in the condition~(\ref{lastcon})---   to a more symmetric  index-eight `projection',  $P_{abcd}{}^{efgh}P_{efgh}{}^{klmn}=P_{abcd}{}^{klmn}$, as in the case of T-geometry~\cite{Jeon:2011cn}.  However, such modification would better not ruin the nice properties of $H_{abcd}$  as (\ref{deltaXG}), (\ref{Hab}), (\ref{Habc}), (\ref{Hbb}). } In fact, generically,  
\be
D_{A}T^{B_{1}\cdots B_{m}}{}_{C_{1}\cdots C_{n}}\neq\na_{A}T^{B_{1}\cdots B_{m}}{}_{C_{1}\cdots C_{n}}\,.
\ee
In any case, the alternative  semi-covariant derivative is compatible with the $\SL(10)$ generalized metric,
\be
\ba{ll}
D_{A}M_{BC}=0\,,~~~~&~~~~D_{A}M^{BC}=0\,,
\ea
\ee
and furthermore, the new connection, $A_{AB}{}^{C}$~(\ref{Acon}), is the \textit{unique} connection which satisfies the above compatibility condition and the symmetric property, $A_{ABC}=A_{ACB}$,  \textit{c.f.~}(\ref{Asym}).\\

The commutator of the above semi-covariant derivatives (\ref{semicov2}) has the expression, 
\be
\ba{l}
{}\left[D_{A},D_{B}\right]T^{C_{1}\cdots C_{m}}{}_{D_{1}\cdots D_{n}}\\
\!\!=\textstyle{\frac{1}{8}} (m-n) R_{ABE}{}^{E}T^{C_{1}\cdots C_{m}}{}_{D_{1}\cdots D_{n}}
-\sum_{i}T^{C_{1}\cdots E\cdots C_{m}}{}_{D_{1}\cdots D_{n}}R_{ABE}{}^{C_{i}}
+\sum_{j}R_{ABD_{j}}{}^{E}T^{C_{1}\cdots C_{m}}{}_{D_{1}\cdots E\cdots D_{n}}\\
\quad +\left(A_{AB}{}^{E}-A_{BA}{}^{E}-\textstyle{\frac{1}{8}}A_{AF}{}^{F}\delta_{B}{}^{E}+
\textstyle{\frac{1}{8}}A_{BF}{}^{F}\delta_{A}{}^{E}\right)D_{E}T^{C_{1}\cdots C_{m}}{}_{D_{1}\cdots D_{n}}\,,
\ea
\ee
where $R_{ABC}{}^{D}$ denotes the standard field strength of the   connection,
\be
\ba{ll}
R_{ABC}{}^{D}&\!:=\partial_{A}A_{BC}{}^{D}-\partial_{B}A_{AC}{}^{D}+A_{AC}{}^{E}A_{BE}{}^{D}-A_{BC}{}^{E}A_{AE}{}^{D}\\
{}&=D_{A}A_{BC}{}^{D}+\textstyle{\frac{1}{8}}A_{AE}{}^{E}A_{BC}{}^{D}-A_{AB}{}^{E}A_{EC}{}^{D}+A_{BC}{}^{E}A_{AE}{}^{D}\,-\,(A\leftrightarrow B)\,.
\ea
\ee
Arbitrary variations of the metric, $\delta M_{AB}$, induces
\be
\ba{rl}
\delta A_{ABC}=&\delta(A_{AB}{}^{D}M_{DC})=\quarter M_{BC}M^{DE}D_{A}\delta M_{DE}-\half D_{A}\delta M_{BC}+A_{A(B}{}^{D}\delta M_{C)D}\,,\\
\delta R_{ABCD}=& D_{A}\delta A_{BCD}-A_{BC}{}^{E}D_{A}\delta M_{ED}+A_{BC}{}^{E}A_{AE}{}^{F}\delta M_{FD}\\
{}&+\textstyle{\frac{1}{8}} A_{E}{}^{E}\left(\quarter M_{CD} M^{FG}D_{B}\delta M_{FG}-\half D_{B}\delta M_{CD}+A_{B(C}{}^{F}\delta M_{D)F}\right)\\
{}&-A_{AB}{}^{E}\left(\quarter M_{CD} M^{FG}D_{E}\delta M_{FG}-\half D_{E}\delta M_{CD}+A_{E(C}{}^{F}\delta M_{D)F}\right)\\
{}&\,-\,(A\leftrightarrow B)\,,
\ea
\ee
which may be useful to address a higher dimensional U-geometry in future.\\

\section{Useful formulae}
The generalized Lie derivative and the semi-covariant derivative of Kronecker delta symbol are all  trivial,
\be
\ba{ll}
\hcL_{X}\delta^{a}_{~b}=0\,,\quad&\quad\na_{cd}\delta^{a}_{~b}=0\,.
\ea
\ee
For a generic covariant tensor density satisfying (\ref{covT}), using (\ref{semicovT}),  we have
\be
\ba{l}
\delta_{X}(\na_{ab}\na_{cd}T^{e_{1}e_{2}\cdots e_{p}}{}_{f_{1}f_{2}\cdots f_{q}})\\
=\hcL_{X}\left(\na_{ab}\na_{cd}
T^{e_{1}e_{2}\cdots e_{p}}{}_{f_{1}f_{2}\cdots f_{q}}\right)\\
\quad~-\quarter\sum_{i=1}^{p}\left(T^{e_{1}\cdots g\cdots e_{p}}{}_{f_{1}\cdots f_{q}}\na_{ab}H_{cdg}{}^{e_{i}}+\na_{ab}T^{e_{1}\cdots g\cdots e_{p}}{}_{f_{1}\cdots f_{q}}H_{cdg}{}^{e_{i}}
+\na_{cd}T^{e_{1}\cdots g\cdots e_{p}}{}_{f_{1}\cdots f_{q}}H_{abg}{}^{e_{i}}\right)\\
\quad~+\quarter\sum_{j=1}^{q}\left(\na_{ab}H_{cdf_{j}}{}^{g}T^{e_{1}\cdots e_{p}}{}_{f_{1}\cdots g\cdots f_{q}}+H_{abf_{j}}{}^{g}\na_{cd}T^{e_{1}\cdots e_{p}}{}_{f_{1}\cdots g\cdots f_{q}}+H_{cdf_{j}}{}^{g}\na_{ab}T^{e_{1}\cdots e_{p}}{}_{f_{1}\cdots g\cdots f_{q}}\right)\\
\quad~+\quarter H_{abc}{}^{g}\na_{gd}T^{e_{1}\cdots e_{p}}{}_{f_{1}\cdots f_{q}}
+\quarter H_{abd}{}^{g}\na_{cg}T^{e_{1}\cdots e_{p}}{}_{f_{1}\cdots f_{q}}\,.
\ea
\label{semicov2T}
\ee
From $\,J^{almnefgh}H_{blmn}\Gamma_{efgh}=0\,$ (\ref{lastcon}),\,   we have
\be
H_{a}{}^{lmn}(\Gamma_{blmn}-\half\Gamma_{mnlb}+\half\Gamma_{mnbl})
-\textstyle{\frac{1}{3}}(H_{ambn}+H_{bman})\Gamma_{l}{}^{mln}=0\,.
\ee
Contracting free indices, $a,b$, and from (\ref{Hab}), we note
\be
H^{abcd}\Gamma_{abcd}=0\,.
\label{HG4}
\ee
This further implies with  $H_{[abc]d}\Gamma^{abcd}=0$,
\be
H^{abcd}\Gamma_{acbd}=0\,.
\label{HG42}
\ee
In order to verify (\ref{varscalarR}), we need
\be
\ba{ll}
\left[(\delta_{X}-\hcL_{X})\Gamma_{ca}{}^{c}{}_{b}\right]\Gamma_{d}{}^{bda}\!\!\!\!\!&=\half(\delta_{X}-\hcL_{X})(\Gamma_{ca}{}^{c}{}_{b}\Gamma_{d}{}^{bda})\,,\\
\left[(\delta_{X}-\hcL_{X})\Gamma_{abcd}\right]\Gamma^{bdca}\!\!\!\!\!&=\left[(\delta_{X}-\hcL_{X})\Gamma_{bdca}\right]\Gamma^{dacb}=\left[(\delta_{X}-\hcL_{X})\Gamma_{bdca}\right]\Gamma^{abcd}\\
{}&=\half(\delta_{X}-\hcL_{X})(\Gamma_{abcd}\Gamma^{bdca})\,,\\
\left[(\delta_{X}-\hcL_{X})\Gamma_{abcd}\right]\Gamma^{bdac}\!\!&=
\left[(\delta_{X}-\hcL_{X})\Gamma_{abcd}\right]\Gamma^{dabc}=
\half\left[(\delta_{X}-\hcL_{X})\Gamma_{abcd}\right](\Gamma^{bdac}+\Gamma^{dabc})\\
{}\!\!\!\!\!&=-\half\left[(\delta_{X}-\hcL_{X})\Gamma_{abcd}\right]\Gamma^{abdc}=-\quarter(\delta_{X}-\hcL_{X})(\Gamma_{abcd}\Gamma^{abdc})\,,
\ea
\label{usefulGGd}
\ee
and 
\be
\ba{ll}
2\Gamma_{abcd}\Gamma^{cadb}&=\Gamma_{abcd}(\Gamma^{cadb}+\Gamma^{bcda})=\Gamma_{bcad}\Gamma^{cadb}+\Gamma_{cabd}\Gamma^{bcda}\\
{}&=\half(\Gamma_{abcd}+\Gamma_{bcad})\Gamma^{cadb}+\half(\Gamma_{abcd}+\Gamma_{cabd})\Gamma^{bcda}=-\half\Gamma_{cabd}\Gamma^{cadb}-\half\Gamma_{bcad}\Gamma^{bcda}\\
{}&=-\Gamma_{abcd}\Gamma^{abdc}\,.
\ea
\label{usefulGG2to1}
\ee

\end{document}